\documentclass[12pt,preprint]{aastex}

\usepackage{graphics}

\newcommand{\be}{\begin{equation}}
\newcommand{\ee}{\end{equation}}

\slugcomment{submitted to : ApJ }

\shorttitle{BH masses and Doppler factors of $\gamma$-ray AGNs}
\shortauthors{Fan Z.H.}

\begin{document}

\title{Black hole masses and Doppler factors of \\
$\gamma$-ray active galactic nuclei}

\author{Zhong-Hui Fan\thanks{Email: fanzh@center.shao.ac.cn} , and
Xinwu Cao\thanks{Email: cxw@center.shao.ac.cn}} \affil{Shanghai
Astronomical Observatory, Chinese Academy of Sciences, Shanghai,
200030, China}

\begin{abstract}
The sizes of the broad-line regions (BLRs) in $\gamma$-ray active
galactic nuclei (AGNs) are estimated from their optical continuum
luminosity by using the empirical relation between $R_{\rm BLR}$
and $L_{\lambda,\rm opt}$. Using the broad emission line data, we
derive the photon energy density in the relativistic blobs near
the massive black holes in AGNs. We calculate the power of the
broad-line photons Compton up-scattered by the relativistic
electrons in the blobs. Compared with observed $\gamma$-ray
emission data, the Doppler factors $\delta$ of the blobs for a
sample of 36 $\gamma$-ray AGNs are derived, which are in the range
of $\sim 3-17$. We estimate the central black hole masses of these
$\gamma$-ray AGNs from the sizes of the BLR and the broad line
widths. It is found that the black hole masses are in the range of
$\sim 10^{8}-10^{10}M_{\odot}$. A significant correlation is found
between the Doppler factor $\delta$ and the core dominance
parameter $R$. The results are consistent with the external
radiation Compton (ERC) models for $\gamma$-ray emission from
AGNs. The soft seed photons are probably from the broad-line
regions.
\end{abstract}


\section{Introduction}
All $\gamma$-ray AGNs are identified as flat-spectrum radio
sources. The third catalog of high-energy $\gamma$-ray sources
detected by the Energetic Gamma Ray Experiment Telescope (EGRET)
on the Compton Gamma Ray Observatory (CGRO) includes 66
high-confidence identifications of blazars and 27 lower confidence
potential blazar identifications \citep{H99}. This offers a good
sample for the explorations on the radiative mechanisms of
$\gamma$-rays from AGNs. The violent variations in very short
time-scales imply that the $\gamma$-ray emission is closely
related with the relativistic jets in blazars. There are two kinds
of models, namely, leptonic models and hadronic models, proposed
for $\gamma$-ray emission in blazars (see Mukherjee 2001 for a
review). According to the different origins of the soft photons,
the leptonic models can be classified as two groups: synchrotron
self-Compton (SSC) models and ERC models (see Sikora $\&$ Madejski
2001 for a review).

In the frame of SSC models, the synchrotron photons are both
produced and Compton up-scattered by the same population of
relativistic electrons in the jets of $\gamma$-ray blazars. The
synchrotron radiation is responsible for the low energy component
in radio bands, and the synchrotron photons are Compton
up-scattered to $\gamma$-ray photons by the same population of
relativistic electrons in the jets \citep{M92}. However, SSC
models meet difficulties with the observed rapidly variable fluxes
in MeV-GeV for some blazars. It has been realized that the
processes other than SSC may occur at least in some $\gamma$-ray
blazars. One possibility is soft seed photons being from the
external radiation fields outside the jets, namely, ERC models.
The origins of soft seed photons may include the cosmic microwave
background radiation, the radiation of the accretion disk
(including photons from the disk scattered by surrounding gas and
dust), infrared emission from the dust or/and a putative molecular
torus, and broad-line regions, etc \citep{D02}. Recently,
\citet{S02} proposed that the external radiation is from the BLRs
for GeV $\gamma$-ray blazars with flat $\gamma$-ray spectra, while
the near-IR radiation from the hot dust is responsible for MeV
$\gamma$-ray blazars with steep $\gamma$-ray spectra. In their
model, the electrons are assumed to be accelerated via a two-step
process and their injection function takes the form of a double
power law with a break at the energy that divides the regimes for
two different electron acceleration mechanisms.

The Doppler boosting factor $\delta$ of the jets is a crucial
quantity in studying the physical processes at work in the regions
near the massive black holes in AGNs. Several different approaches
are proposed to estimate the Doppler factors $\delta$ of the jets
in AGNs. \citet{G93} derived the synchrotron self-Compton Doppler
factor $\delta_{\rm SSC}$ of the jets in AGNs from the VLBI core
sizes and fluxes, and X-ray fluxes, on the assumption of the X-ray
emission being produced by the SSC processes in the jets.
\citet{GD96} assumed energy equipartition between the particles
and the magnetic fields in the jet components, and derived the
equipartition Doppler factor $\delta_{\rm eq}$. More recently, the
variability Doppler factor $\delta_{\rm var}$ is derived on the
assumption that the associated variability brightness temperature
of total radio flux density flares are caused by the relativistic
jets \citep{L99}.

In this paper, we use broad-line and $\gamma$-ray emission data to
derive the Doppler boosting factors $\delta_{\gamma}$ for a sample
of $\gamma$-ray blazars in the frame of the ERC model. The
cosmological parameters $H_{0}$=75 km s$^{-1}$Mpc$^{-1}$ and
$q_{0}$=0.5 have been adopted in this work.

\section{Model}
\subsection{Inverse Compton radiative processes}
In this work, we calculate the power of $\gamma$-ray emission from
blazars by the external Compton processes mainly following the
calculations of \citet{D97}. We then extend it to the case of the
soft seed photons being from the BLRs of the blazars.

Assuming that blobs are moving with a constant velocity at an
angle $\mu_{\rm obs}$=cos $i$ with respect to the line of sight,
we have the inverse Compton scattered $\gamma$-ray flux density
$S_{\rm C}$ \citep{D97}:
\begin{equation}
S_{\rm
C}={\frac{\delta_{\gamma}^{4+2\alpha}}{\alpha+2}}{\frac{c\sigma_{\rm
T}u_{\rm i}^{*}n_{\rm eo}V_{\rm b}} {16\pi d_{\rm
L}^{2}}}(1+z)^{1-\alpha}{\frac{(1+\mu_{\rm
obs})^{2+\alpha}}{\mu_{\rm obs}}} {\frac{\epsilon_{\rm
obs}^{-\alpha}}{(\overline{\epsilon}^{*})^{1-\alpha}}},
\end{equation}
where $\delta_{\gamma}$ is the Doppler factor, $\alpha$ is the
$\gamma$-ray photon energy spectral index, $z$ is the cosmological
redshift, $u_{\rm i}^{*}$ is the energy density of the soft seed
photons in the blob's frame in units of ergs cm$^{-3}$, $d_{\rm
L}$ is the luminosity distance, $\overline{\epsilon}^{*}$ is the
dimensionless monochromatic mean soft photon energy in units of
the electron rest mass energy, $n_{\rm eo}$ is the number density
of electron in the blob, $V_{\rm b}$ is the volume of the blob,
and $\sigma_{\rm T}$ is the Thomson cross section.

An optically thin, homogenous spherical blob is assumed to be
located very near the central black hole, so most of the soft seed
photons from the BLRs only experience one or a few inverse Compton
scatters in the blob. The total cross section $S_{\rm T}\simeq
\sigma_{\rm T}n_{\rm eo}V_{\rm b}=\xi \pi r_{\rm b}^{2}$, where
$r_{\rm b}$ is the radius of the blob. For the homogenous
spherical blob, we have $\xi =4\tau /3$, where $\tau = \sigma
_{\rm T}n_{\rm eo}r_{\rm b}$ is the optical depth of the blob. As
we do not know the value of the electron density $n_{\rm eo}$ in
the blob, the parameter $\xi$, which is required $\xi \leq 1$, is
adopted to describe the optical depth of the blob in our
calculations.

For $\gamma$-ray blazars, the jets are believed to move in the
direction very close to the line of sight, so $\mu_{\rm obs}\simeq
1$ is taken and Eq. (1) becomes
\begin{equation}
S_{\rm
C}={\frac{\delta_{\gamma}^{4+2\alpha}}{\alpha+2}}{\frac{c\xi
r_{\rm b}^{2}u_{\rm i}^{*}}{16 \pi d_{\rm L}^{2}}}
(1+z)^{1-\alpha}2^{2+\alpha}{\frac{\epsilon_{\rm obs}^{-\alpha}}
{(\overline{\epsilon}^{*})^{1-\alpha}}} .
\end{equation}
The soft seed photons are assumed to be from the different line
emission. The observed $\gamma$-ray flux density is the sum of
inverse Compton scattered flux densities contributed by different
emission lines:
\begin{equation}
S_{\rm C}=S_{\rm C,H\beta}+S_{\rm C,MgII}+S_{\rm C,CIV}+S_{\rm
C,Ly\alpha} \cdots = \sum_{\rm i}^{\rm n} S_{\rm C,i} .
\end{equation}

\subsection{Size of the blob}
The size of the blob can be estimated from the observed
$\gamma$-ray variability time-scale. \citet{H96} derived the size
of the blobs $\sim100R_{\rm g}$ ($R_{\rm g}=2GM_{\rm bh}/c^{2}$)
from the time-scale for substantial $\gamma$-ray variations of
3C279, if a $10^{8} M_{\odot}$ black hole is present in 3C279.
\citet{GM96} derived the upper limit on the blob-to-disc distance
from the observed variability timescale $t_{\rm var}$ of the
$\gamma$-ray fluxes. As the timescale $t_{\rm var}\sim (r_{\rm
b}/\delta c)$, the radii of the blobs are around
$3\times10^{16}(t_{\rm var}/d)(\delta/10)$ cm. As the typical
variable timescale, $t_{\rm var}\sim1$ day, we can infer the radii
of the blobs are $\sim3\times10^{16}$ cm if the Doppler factor
$\delta \sim 10$ is adopted. For typical radio-loud quasars having
masses of black holes around $10^{9}M_{\odot}$, so the radii
$r_{\rm b}$ of the blobs are around $200R_{\rm g}$. \citet{M97}
employed the ERC model to explain $\gamma$-ray emission from the
blazars. The comparisons of their model with the observed spectrum
of 3C273 give the radius of the blob $r_{\rm b}\sim
2\times10^{16}$ cm. The central black hole mass of 3C273 is around
$5.5\times 10^{8}M_{\odot}$ \citep{K00}. Thus, we get $r_ {\rm
b}\sim 123R_{\rm g}$ for the blob in 3C273. In this paper, we take
$r_{\rm b}=120R_{\rm g}$ in our calculations.

\subsection{Soft photon energy density in the blob's flame}
The observed flux of the broad emission line is
\begin{equation}
f_{\rm line,i} = \frac{L_{\rm BLR,i}} {4\pi d_{\rm L}^{2}} ,
\end{equation}
where $L_{\rm BLR,i}$ is the luminosity of a single broad emission
line.

The energy density of this emission line in the blob is given by
\begin{equation}
u_{\rm i}^{*}={\frac{1}{c}}\left(\frac{d_{\rm L}}{R_{\rm
BLR,i}}\right)^{2}f_{\rm line,i} ,
\end{equation}
where $u_{\rm i}^{*}$ is monochromatic seed photon energy density
of the broad line in the blobs and $R_{\rm BLR,i}$ is the radius
of the BLR. The total soft seed photon energy density $u_{\rm
tot}^{*}=\sum u_{\rm i}^{*}$ is the sum of all emission lines. Eq.
(5) is only valid for the blob near the central black hole. At
further distance away from the central black hole, the energy
density of BLR photons in the blob is anisotropic and can be
treated analogously to the approach for the dust torus
\citep{A02}. In this paper, however, we only consider the case of
the blob being located near the central black hole.

\subsection{Sizes of broad emission line regions}
Based on the long-time observations of 17 nearby Seyfert 1
galaxies and 17 Palomar-Green quasars, \citet{K00} found an
empirical relation between the size of the BLR $R_{\rm BLR}$ of
H$\beta$ and the monochromatic continuum luminosity at 5100 $\rm
\AA$:
\begin{equation}
R_{\rm BLR}=32.9 \times \left(\frac{\lambda L_{5100}}{10^{37} \rm
W}\right)^{0.7} {\rm lt-day} .
\end{equation}

\citet{M02} suggested that Mg\,{\sc ii} is a low-ionization line
as H$\beta$, so that Mg\,{\sc ii} is expected to be produced in
the same region as H$\beta$. For a sample of 22 objects, they
found a correlation between the FWHM of Mg\,{\sc ii}  and
H$\beta$: $V_{\rm FWHM,MgII}\sim V_{\rm FWHM,H\beta}^{1.02}$. It
is therefore reasonable to expect that Mg\,{\sc ii} and H$\beta$
are produced in the same region. Using the same sample and the
size of the BLR derived by \citet{K00}, they obtained a
correlation between $R_{\rm BLR}$ and the monochromatic continuum
luminosity at 3000 $\rm \AA$:
\begin{equation}
R_{\rm BLR}=28.4 \times \left(\frac{\lambda L_{3000}}{10^{37} \rm
W}\right)^{0.47} {\rm lt-day} ,
\end{equation}
which has been corrected to the cosmological parameters used in
this paper.

\citet{V02} suggested a relation between the black hole mass
$M_{\rm bh,UV}$ and the monochromatic continuum luminosity at 1350
$\rm \AA$. The black hole mass $M_{\rm bh,UV}$ is the central
black hole mass based on the calibration of the single-epoch UV
spectral measurements. They derived
\begin{equation}
M_{\rm bh,UV}=1.6\times10^{6} \left(\frac{V_{\rm FWHM,CIV}}{10^{3}
\rm km s^{-1}}\right)^{2}\left(\frac{\lambda L_{1350}}{10^{37}{\rm
W}}\right)^{0.7}M_{\odot} ,
\end{equation}
where $V_{\rm FWHM}$ is the full width at half-maximum of the line
C\,{\sc iv}. The relation between $R_{\rm BLR}$ and the
monochromatic continuum luminosity at 1350 $\rm \AA$ is therefore
available:
\begin{equation}
R_{\rm BLR}=10.9 \times \left(\frac{\lambda L_{1350}}{10^{37} \rm
W}\right)^{0.7} {\rm lt-day} .
\end{equation}

Now, we can derive the sizes of broad emission lines H$\beta$,
Mg\,{\sc ii}, and C\,{\sc iv} from the optical or UV continuum
luminosity by using the empirical relations (6), (7) and (9),
respectively.

\subsection{The black hole mass}
We can estimate the central black hole masses of AGNs on the
assumption that the motions of the gases in BLRs are virilized
from the width of the emission line, if the size of the BLR is
available. We use the empirical relation between the size of the
BLR $R_{\rm BLR}$ and the optical continuum luminosity
$L_{\lambda}$ given in last subsection to estimate the size of the
BLR $R_{\rm BLR}$. The black hole mass is given by
\begin{equation}
M_{\rm bh}=2.25 \times R_{\rm BLR}V_{\rm FWHM}^{2}G^{-1} ,
\end{equation}
if the FWHM of any broad-line of H$\beta$, Mg\,{\sc ii}, and
C\,{\sc iv} is available \citep{G01}.

\subsection{The Doppler factor $\delta_{\gamma}$}
The Doppler factor $\delta_{\gamma}$ is given by
\begin{equation}
\delta_{\gamma}=2^{{\frac{2-\alpha}{4+2\alpha}}}
(1+z)^{{\frac{\alpha-1}{4+2\alpha}}}(\alpha+2)^{{\frac{1}{4+2\alpha}}}
\left({\frac{c\xi r_{\rm b}^{2}} {S_{\rm C} d_{\rm
L}^{2}}}\right)^{{-\frac{1}{4+2\alpha}}} \left(\sum \frac{u_{\rm
i}^{*}}{(\overline{\epsilon}_{\rm
i}^{*})^{1-\alpha}}\right)^{{-\frac{1}{4+2\alpha}}} \epsilon_{\rm
obs}^{{\frac{\alpha}{4+2\alpha}}} .
\end{equation}
Using the observed $\gamma$-ray flux, the Doppler factor
$\delta_{\gamma}$ is available from Eq. (11), if the soft seed
photon energy density $u_{\rm i}^{*}$ in the blob, the radius of
the blob $r_{\rm b}$, and $\xi$ are available. As the parameter
$\xi$ is required to be less than unit, we can obtain the lower
limit of $\delta_{\gamma}$, if $\xi=1$ is adopted.

\section{The sample}
We search the literature and collect all available data of broad
emission lines of $\gamma$-ray AGNs in EGRET catalog III. The
selection criterion is that the sources of which the FWHM of at
least one of the following broad emission lines H$\beta$, Mg\,{\sc
ii} and C\,{\sc iv} and the fluxes are available. This leads to 36
sources, of which 30 sources are the high-confidence
identification blazars and 6 sources are the lower confidence
potential  blazar identifications listed in \citet{H99}. For the
source 0954$+$658, only the FWHM and the flux of the broad line
H$\alpha$ are available. We estimate the BLR size of H$\alpha$
using the empirical relation $R_{\rm BLR}(\rm H\alpha)=1.19 R_{\rm
BLR}(\rm H\beta)$ proposed by \citet{K00}.

In principle, all broad emission line fluxes are needed to
calculate the total soft seed photon energy in the blobs
contributed by different broad emission lines. Usually, the fluxes
of only one or several broad emission lines are available for most
sources in our sample due to the restriction of redshift. In our
calculations, we consider fluxes for the following lines:
H$\beta$, Mg\,{\sc ii}, C\,{\sc iv}, C\,{\sc iii}, Ly$\alpha$ and
H$\alpha$, which contribute about 60 per cent of the total broad
line emission. We use the line ratios presented by \citet{F91}, in
which the relative strength of Ly$\alpha$ is taken as 100, to
calculate the total broad line emissions. \citet{CP97} added the
contribution from the flux of H$\alpha$ $F_{\rm H\alpha}$, with a
value of 77. This gives a total relative flux $<F_{\rm
BLR}>=536.04$ (narrow lines are not included). For the lines
H$\beta$, Mg\,{\sc ii} and C\,{\sc iv}, we use Eq. (6), (7) and
(9) to estimate their radii of the BLRs. For other lines of which
the FWHM is available, we compare their FWHM with that of any one
of those three lines H$\beta$, Mg\,{\sc ii} and C\,{\sc iv} to
estimate the radii of the BLRs. We use Eq. (5) to calculate the
soft photon energy density in the blob for the lines with
available FWHM and flux data. We calculate the total soft seed
photon energy density $u_{\rm tot}^{*}$ in the blob as
\begin{equation}
u_{\rm tot}^{*}=\sum u_{\rm i}^{*}=536.04 \times {\frac{\sum_{\rm
i}u_{\rm i, obs}^{*}}{\sum_{\rm i}F_{\rm i, re}} },
\end{equation}
where $\sum_{\rm i}u_{\rm i,obs}^{*}$ is the photon density in the
blob contributed by the lines with available emission line data,
$\sum_{\rm i}F_{\rm i, re}$ is the sum of relative fluxes of these
lines.

\section{Results}
Using the method described in $\S$ 2, we derive the parameters for
all the sources in our sample. All the observational data and the
results are listed in Tables $1-3$. The $\gamma$-ray emission is
violently variable. For each source, we derived two values of the
Doppler factor $\delta_{\gamma}$ corresponding to the lowest and
highest values of the observed $\gamma$-ray fluxes.

In Figs. \ref{1} and \ref{2}, we compare the Doppler factors
$\delta_{\gamma}$ for the lowest and highest $\gamma$-ray flux
cases with $\delta_{\rm var}$ derived from the variability
time-scale of radio emission, respectively \citep{L99}. It is
found that the values of $\delta_{\rm var}$ are higher than that
of $\delta_{\gamma}$ for most sources in our sample.

\begin{figure}
\plotone{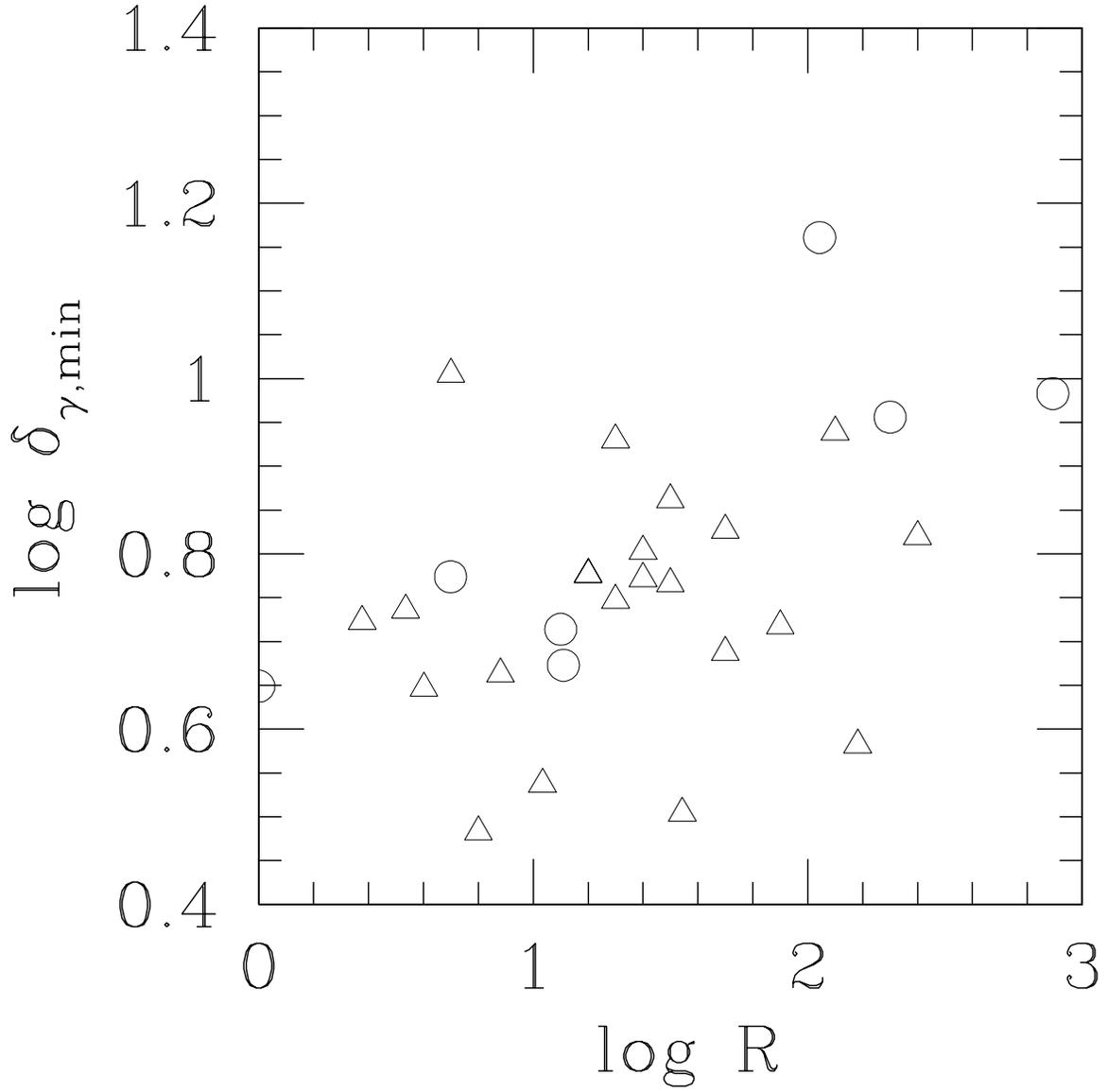} \caption{The core dominance parameter $R$ vs. the
Doppler factor $\delta_{\gamma,\rm min}$. \label{3}}
\end{figure}

\begin{figure}
\plotone{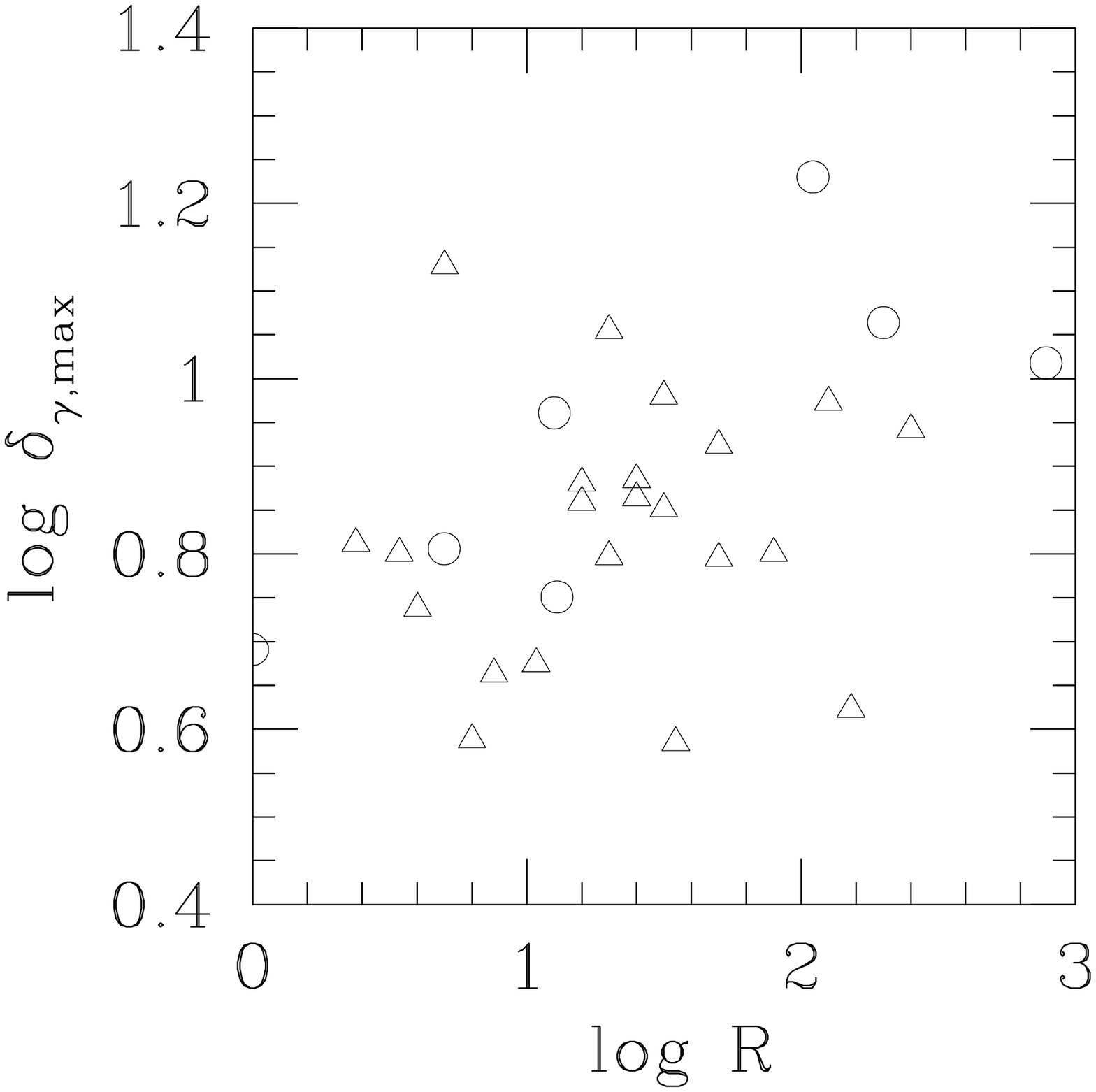} \caption{Same as Fig. 3, but the Doppler factor
$\delta_{\gamma,\rm max}$ rederived from the highest $\gamma$-ray
fluxes. \label{4}}
\end{figure}

In Fig. \ref{3}, we plot the relation between the core dominance
parameter $R$ and the Doppler factor $\delta_{\gamma, \rm min}$
derived from the lowest $\gamma$-ray flux. The core dominance
parameters $R$ are taken from \citet{C01}, except those of the
sources of 0454$-$234, 0521$-$365 and 1741$-$038 taken from
\citet{D95}. A correlation between $R$ and $\delta_{\gamma, \rm
min}$ is found at a significant level of 97.5 per cent (Spearman
correlation coefficient $\rho$). The results are plotted in Fig.
\ref{4} for the Doppler factor $\delta_{\gamma, \rm max}$ derived
from the highest $\gamma$-ray flux. The significant level of the
correlation becomes 95.2 per cent.

\begin{figure}
\plotone{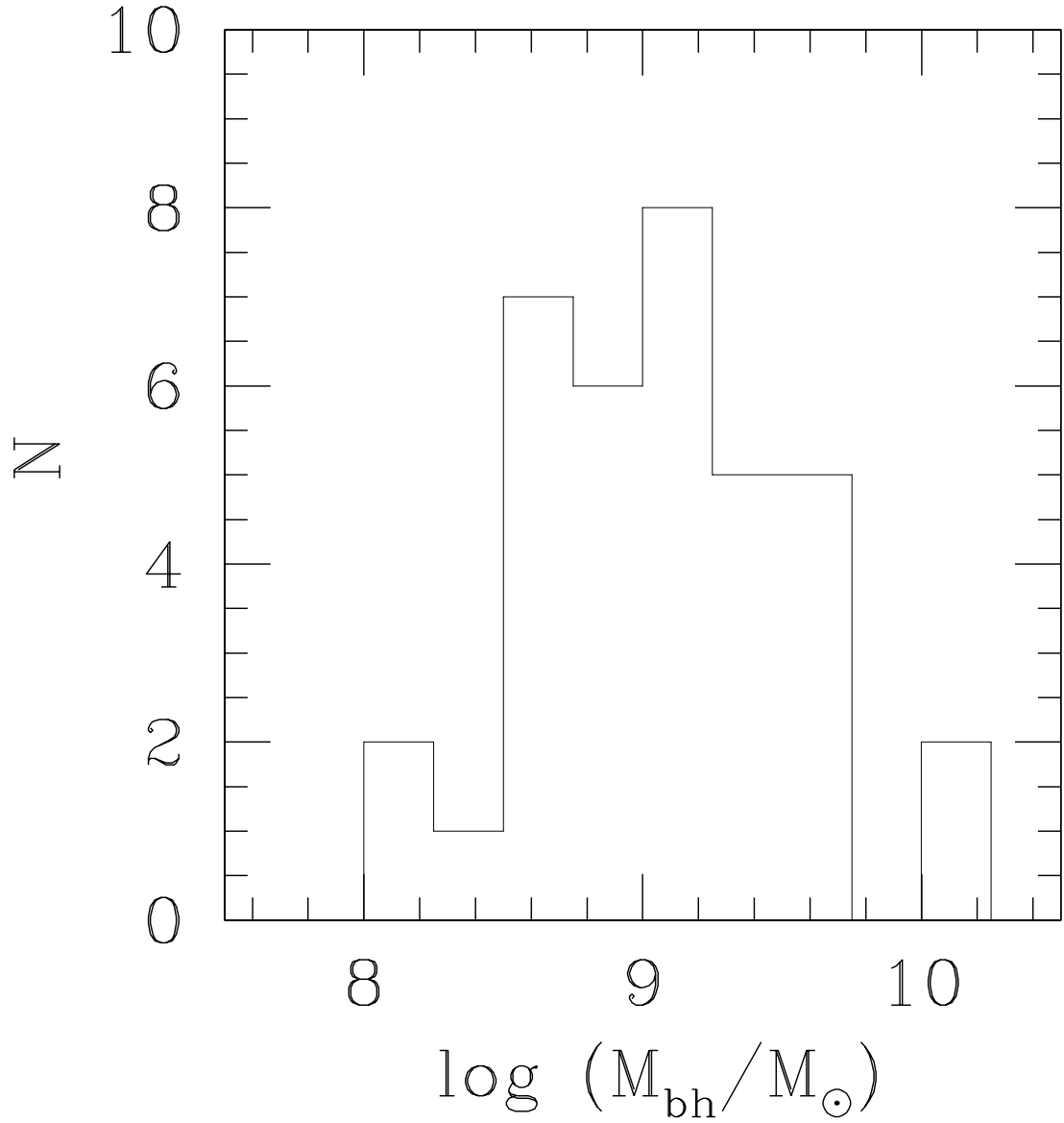} \caption{The distribution of the black hole
masses. \label{5}}
\end{figure}

The distribution of the central black hole mass $M_{\rm bh}$ in
$\gamma$-ray blazars is plotted in Fig. \ref{5}. The black hole
masses $M_{\rm bh}$ are in the range of $10^{8} - 10^{10}
M_{\odot}$ with an average of around $2\times10^{9} M_{\odot}$.

\begin{figure}
\plotone{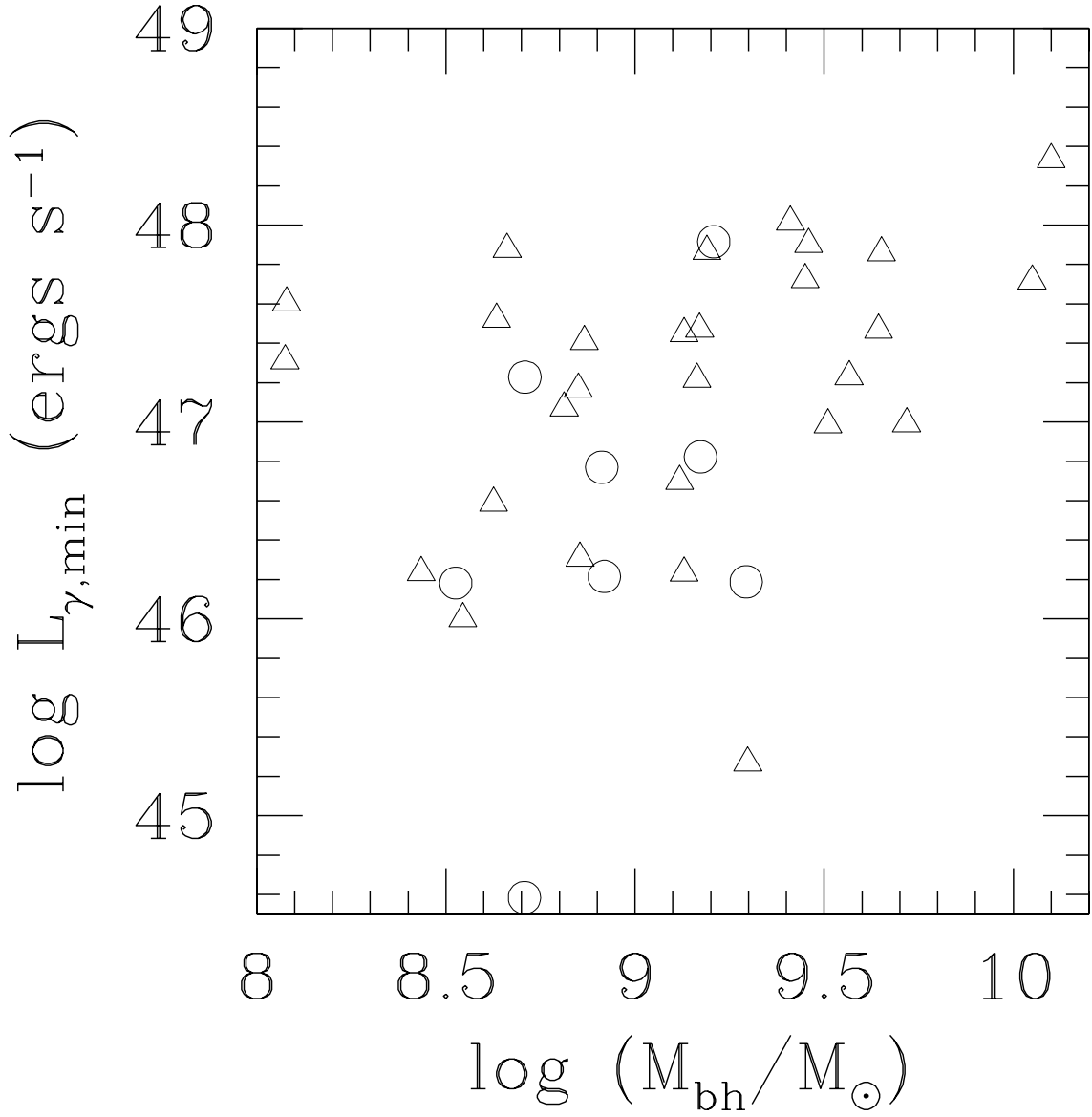} \caption{The lowest $\gamma$-ray luminosity
L$_{\gamma, \rm min}$ vs. the black hole mass $M_{\rm bh}$.
\label{6}}
\end{figure}

\begin{figure}
\plotone{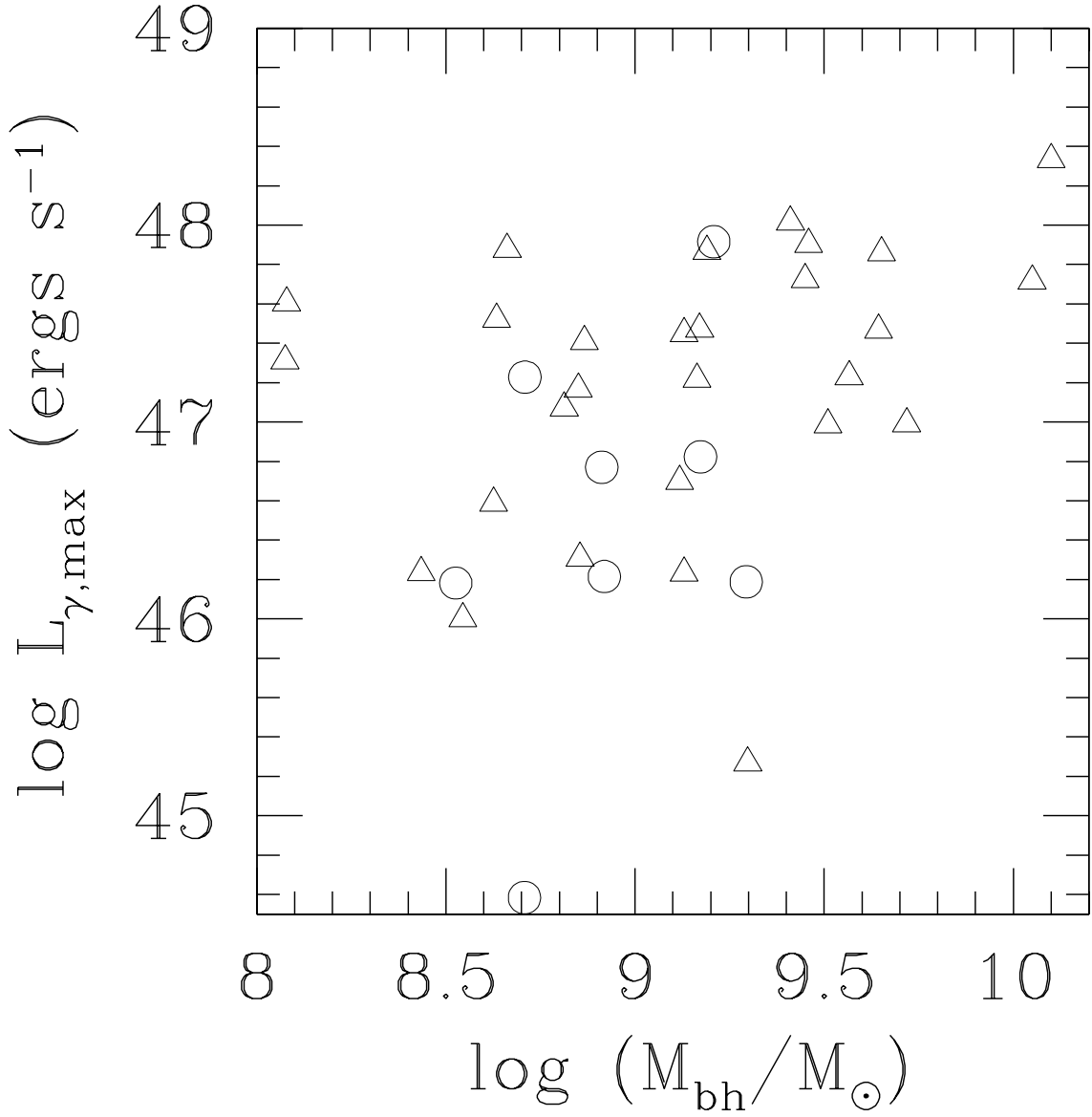} \caption{Same as Fig. 6, but for the highest
$\gamma$-ray luminosity L$_{\gamma, \rm max}$. \label{7}}
\end{figure}

We plot the relation of the black hole mass $M_{\rm bh}$ with the
lowest and highest $\gamma$-ray luminosity $L_{\gamma}$ in Figs.
\ref{6} and \ref{7}, respectively. The correlations between them
are found at the significant levels of 97.8 and 96.6 per cent for
the lowest and highest $\gamma$-ray emissions, respectively.

\section{Discussion}
In this work, we estimate the size of the BLR from the optical or
UV continuum luminosity. The observed optical/UV continuum
emission from $\gamma$-ray blazars is a mixture of the emission
from the disks and jets. The optical/UV continuum emission may be
strongly beamed to us, since the viewing angles of the
relativistic jets in these blazars are small. The emission from
the jets may dominate over that from the disks at least in some
blazars. So, the BLR size derived from the observed optical/UV
continuum luminosity may be over estimated, and the derived black
hole mass is a upper limit \citep{G01}.

The size of the blob $r_{\rm b}$ is assumed to be 120$R_{\rm g}$
in our calculations (see discussion in Section $\S$ 2.2). As the
black hole mass $M_{\rm bh}$ is derived from the size of the BLR
and the line width, the size of the blob $r_{\rm b}$ is
proportional to the BLR size $R_{\rm BLR}$. From Eq. (5), we know
that the energy density in the blob $u^{*} \propto 1/R_{\rm
BLR}^{2}$. The derived Doppler factor $\delta_{\gamma}$ in this
work is therefore independent of the BLR size $R_{\rm BLR}$ (see
Eq. (11)). It indicates that the overestimate of the BLR size
caused by the contamination of the optical/UV continuum emission
has not affected the values of derived Doppler factor
$\delta_{\gamma}$.

In our calculations, we adopted a parameter $\xi$ to describe the
optical depth of the blob, since we do not know the exact value of
the actual number density $n_{\rm eo}$ of the electrons in the
blob. From Eq. (11), it is found that the Doppler factor
$\delta_{\gamma} \propto \xi^{-1/(4+2\alpha)}$. For a typical
value of $\alpha \sim 1.5$, we found that the derived
$\delta_{\gamma}$ becomes about twice of its initial value, if the
parameter $\xi$ is changed from 1 to 0.01. The uncertainty of
$\xi$ in our calculations would not affect our results
significantly.

We only consider the photons from the BLRs as the soft seed
photons. Other sources of soft seed photons (e.g., emission from
the disks, synchrocyclotron emission in the blobs, etc.) have not
been included in the calculations. The Doppler factor
$\delta_{\gamma}$ derived in this work are in the range of 3.04
and 17.00, if $\xi =1$ is adopted. It is generally consistent with
the results derived by other methods \citep{G93,GD96,L99}. It may
imply that the soft seed photons are indeed mainly from the BLRs
in these sources.

\begin{figure}
\plotone{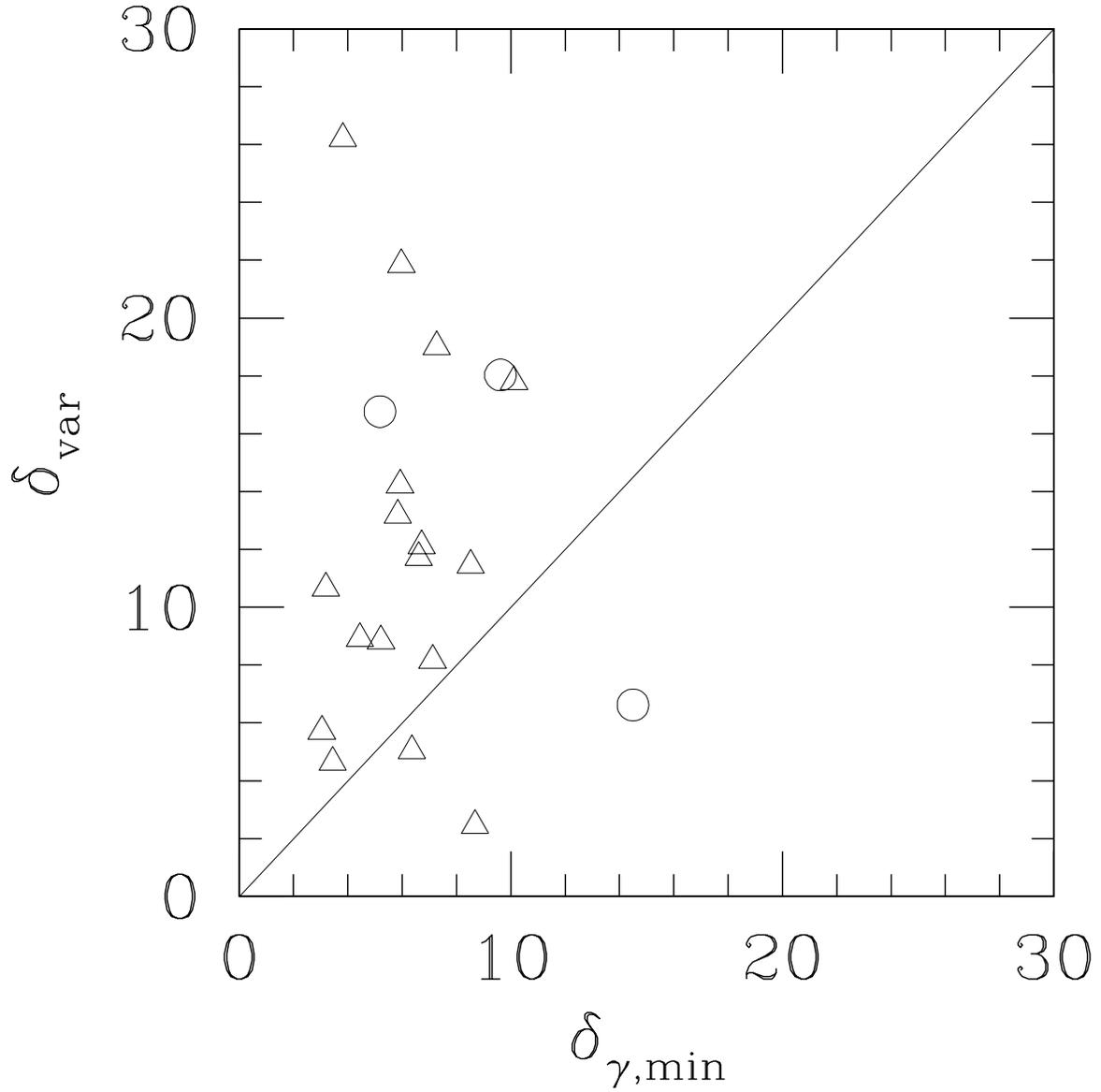} \caption{The variability Doppler factor
$\delta_{\rm var}$ vs. the Doppler factor $\delta_{\gamma,\rm
min}$ derived in this work. The circles represent BL Lac objects,
while the triangles represent quasars. \label{1}}
\end{figure}

\begin{figure}
\plotone{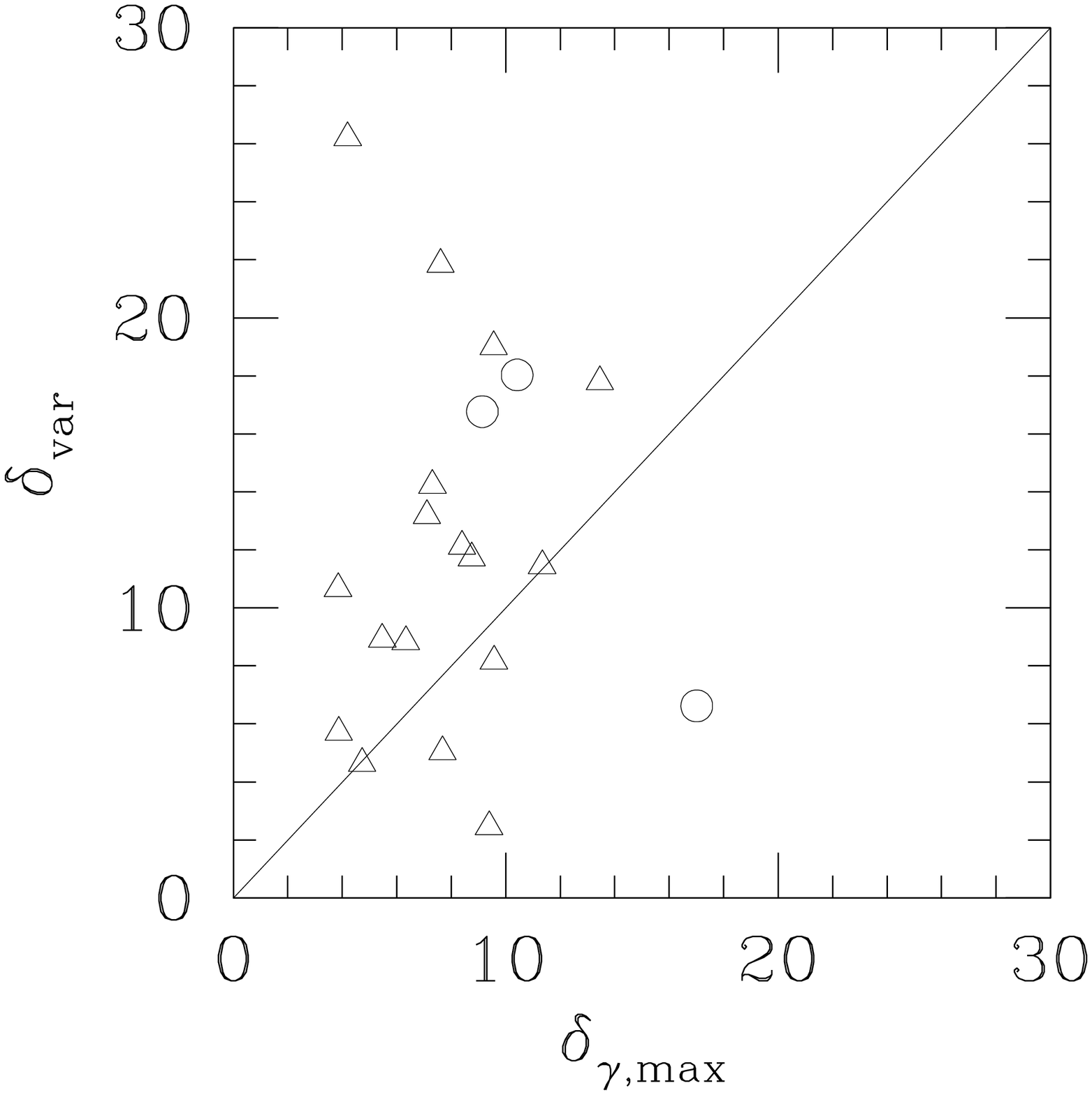} \caption{Same as Fig. 1, but the Doppler factor
$\delta_{\gamma,\rm max}$ rederived from the highest $\gamma$-ray
fluxes. \label{2}}
\end{figure}

The variability Doppler factors $\delta_{\rm var}$ of 20
$\gamma$-ray blazars in our sample have been derived from the
radio variable time-scales are available \citep{L99}. Comparing
with the Doppler factors $\delta_{\gamma}$ derived in this work,
we find that the variability Doppler factor $\delta_{\rm var}$ are
higher than $\delta_{\gamma}$ derived in this work for most
sources (see Figs. \ref{1} and \ref{2}). This may be due to the
fact that we adopt $\xi =1$ and the derived Doppler factor
$\delta_{\gamma}$ are then the lower limits. The value of $\xi$
may possibly be estimated roughly by letting $\delta _{var}=\delta
_{\gamma }$. It can be found that the values of $\xi$ can be low
as $10^{-3}-10^{-2}$ for some sources (see Eq. (11), Figs. \ref{1}
and \ref{2}).

The core dominance parameter $R=f_{\rm c}/f_{\rm e}$, where
$f_{\rm c}$ and $f_{\rm e}$ are the core and extended flux
densities at 5 GHz in the rest frame of the source. The core
dominance parameter $R$ is believed to be a good indicator of the
Doppler factor of the jet. We find a significant correlation
between the core dominance parameter $R$ and the derived Doppler
factor $\delta_{\gamma}$. This implies that the derived Doppler
factors $\delta_{\gamma}$ are reliable for the $\gamma$-ray
blazars in this sample and the soft seed photons are mainly from
the BLRs.

It was argued that the SSC radiative mechanism is dominant in BL
Lac objects \citep{S02}. However, we cannot find significant
differences between BL Lac objects and quasars in our
investigations. The reason may be that the BL Lac objects in our
sample have relative stronger broad line emission compared with
other BL Lac objects listed in EGRET catalog III, since we only
select the $\gamma$-ray blazars with broad-line emission data as
our sample. So these BL Lac objects in our sample are more like
quasars as they have rather strong broad-line emission
\citep{V00}. For these BL Lac objects with relative strong
broad-line emission, the ERC radiative mechanism may probably be
important as that in quasars.

The black hole masses of $\gamma$-ray blazars in this sample are
in the range of $10^{8} - 10^{10}M_{\odot}$. This is consistent
with some previous results for radio-loud AGNs \citep{MD01,L00}.
We have found significant correlations between the black hole mass
$M_{\rm bh}$ and the $\gamma$-ray luminosity $L_{\gamma}$. The
reason may be that the $\gamma$-ray flux depends sensitively on
the Doppler factor $\delta_{\gamma}$ and the values of the Doppler
factor of the $\gamma$-ray AGNs in our sample spread over a large
range ($\sim 3-17$). On the other hand, it is expected that the
$\gamma$-ray flux is independent of the black hole mass in ERC
mechanisms with the soft seed photons being from the BLRs (see Eq.
(2) and discussion in the second paragraph of this section).

\acknowledgments
This work is supported by NSFC(No. 10173016) and
the NKBRSF (No. G1999075403). This research has made use of the
NASA/IPAC Extragalactic Database (NED), which is operated by the
Jet Propulsion Laboratory, California Institute of Technology,
under contract with the National Aeronautic and Space
Administration.

\clearpage

\begin{deluxetable}{cccccccc}
\tabletypesize{\scriptsize} \tablecaption{Broad emission line
data, black hole masses and $\gamma$-ray luminosities}
\tablewidth{0pt} \tablehead{ \colhead{Source} &
\colhead{Redshift}& \colhead{Line} & \colhead{FWHM$^{\rm a}$} &
\colhead{Ref.} & \colhead{log {M$_{\rm bh}$/M$_{\rm \bigodot}$}}
 & \colhead{log {L$_{\rm \gamma,min}$}$^{\rm b}$ } &
\colhead{log {L$_{\rm \gamma,max}$}$^{\rm b}$ }}  \startdata
0119$+$041$^{\rm d}$ & 0.637 & H$\beta$  & 4444 & G01&8.855$^{\rm e}$ & 46.309 & 46.909 \\
0208$-$512$^{\rm c}$ & 1.003 & H$\beta$     & 4070 & W86  & 9.208 & 47.916 & 48.497 \\
0336$-$019 & 0.852 & H$\beta$     & 4875 & G01  & 9.459$^{\rm e}$ & 47.903 & 48.580 \\
0414$-$189 & 1.536 & Mg\,{\sc ii} & 1713 & H78  & 8.075 & 47.310 & 47.996 \\
0420$-$014 & 0.915 & H$\beta$     & 3000 & G01  & 9.510$^{\rm e}$ & 46.987 & 47.826 \\
0440$-$003 & 0.844 & Mg\,{\sc ii} & 3900 & B89  & 8.813 & 47.071 & 47.908 \\
0454$-$234$^{\rm c}$ & 1.009 & Mg\,{\sc ii} & 7891 & S89  & 9.173 & 46.823 & 47.082 \\
0454$-$463 & 0.858 & Mg\,{\sc ii} & 2700 & F83  & 8.626 & 46.587 & 47.204 \\
0458$-$020 & 2.286 & C\,{\sc iv}  & 5200 & B89  & 8.662 & 47.878 & 48.734 \\
0521$-$365$^{\rm c,}$$^{\rm d}$ & 0.055 & H$\beta$  & 6758 & SK93 & 8.708 & 44.584 & 44.889 \\
0537$-$441$^{\rm c}$ & 0.894 & Mg\,{\sc ii} & 3200 & W86  & 8.709 & 47.228 & 47.970 \\
0539$-$057$^{\rm d}$ & 0.839 & Mg\,{\sc ii} & 8598 & S93  & 9.411 & 48.015 & 48.015 \\
0804$+$499$^{\rm d}$ & 1.430 & Mg\,{\sc ii} & 2259 & L96  & 8.634 & 47.522 & 47.781 \\
0836$+$710 & 2.172 & Mg\,{\sc ii} & 3062 & L96  & 9.450 & 47.723 & 48.312 \\
0851$+$202$^{\rm c}$ & 0.306 & H$\beta$     & 3441 & S89  & 8.919 & 46.215 & 46.427 \\
0954$+$556 & 0.901 & C\,{\sc iv}  & 10503 & W95 & 9.718 & 46.990 & 47.851 \\
0954$+$658$^{\rm c}$ & 0.368 & H$\alpha$    & 2084 & L96  & 8.526 & 46.182 & 46.618 \\
1127$-$145$^{\rm d}$ & 1.187 & Mg\,{\sc ii} & 2750 & W86  & 8.850 & 47.168 & 47.964 \\
1222$+$216 & 0.435 & H$\beta$     & 2197 & SM87 & 8.435 & 46.235 & 47.078 \\
1226$+$023 & 0.158 & H$\beta$     & 3416 & K00  & 9.298 & 45.268 & 46.026 \\
1229$-$021 & 1.045 & Mg\,{\sc ii} & 5000 & B89  & 9.118 & 46.698 & 47.199 \\
1253$-$055$^{\rm c}$ & 0.538 & H$\beta$ & 3100 & G01  & 8.912$^{\rm e}$ & 46.770 & 48.229 \\
1331$+$170 & 2.084 & C\,{\sc iv}  & 5740 & C92  & 9.171 & 47.472 & 48.348 \\
1334$-$127$^{\rm c}$ & 0.539 & Mg\,{\sc ii} & 7077 & SK93  & 9.294 & 46.188 & 46.753 \\
1424$-$418 & 1.522 & Mg\,{\sc ii} & 9144 & S89  & 9.652 & 47.860 & 48.418 \\
1504$-$166$^{\rm d}$ & 0.876 & Mg\,{\sc ii} & 5139 & H78  & 9.130 & 47.450 & 47.754 \\
1510$-$089 & 0.361 & H$\beta$     & 3180 & G01  & 9.130$^{\rm e}$ & 46.233 & 46.827 \\
1611$+$343 & 1.401 & H$\beta$     & 5600 & G01  & 10.05$^{\rm e}$ & 47.716 & 48.276 \\
1633$+$382 & 1.814 & Mg\,{\sc ii} & 9657 & L96  & 10.09 & 48.332 & 48.861 \\
1725$+$044 & 0.296 & H$\beta$     & 2090 & G01  & 8.545$^{\rm e}$ & 46.001 & 46.251 \\
1739$+$522 & 1.375 & Mg\,{\sc ii} & 2559 & L96  & 8.866 & 47.410 & 48.076 \\
1741$-$038 & 1.054 & Mg\,{\sc ii} & 11574 & S89 & 9.566 & 47.231 & 47.851 \\
2230$+$114 & 1.037 & C\,{\sc iv}  & 3340 & W95  & 9.164 & 47.217 & 47.847 \\
2251$+$158 & 0.859 & H$\beta$     & 2800 & G01  & 9.644$^{\rm e}$ & 47.467 & 48.141 \\
2320$-$035 & 1.411 & C\,{\sc iv}  & 3300 & B89  & 8.079 & 47.605 & 48.273 \\
2351$+$456 & 1.922 & Mg\,{\sc ii} & 5118 & L96  & 9.190 & 47.867 & 48.427 \\
\enddata
\tablenotetext{a}{in unit of km s$^{-1}$} \tablenotetext{b}{in
unit of ergs s$^{-1}$} \tablenotetext{c}{BL Lac object}
\tablenotetext{d}{lower confidence potential blazar
identifications} \tablenotetext{e}{the data of black hole masses
from \citet{G01}} \tablerefs{B89: \citet{B89}; C92: \citet{C92};
F83: \citet{F83}; G01: \citet{G01}; H78: \citet{H78}; K00:
\citet{K00}; L96: \citet{L96}; S89: \citet{S89}; S93: \citet{S93};
SK93: \citet{SK93}; SM87: \citet{SM87}; W86: \citet{W86}; W95:
\citet{W95}. }
\end{deluxetable}

\clearpage
\begin{deluxetable}{ccccccc}
\tabletypesize{\scriptsize} \tablecaption{Broad emission lines
data and the total energy density of the blob } \tablewidth{0pt}
\tablehead{ \colhead{Source} & \colhead{Lines}&
\colhead{References} & \colhead{log $u_{\rm tot}^{*}$}  }
\startdata
0119$+$041 & H$\beta$ & JB91 & $-$1.896  \\
0208$-$512 & Mg\,{\sc ii}, C\,{\sc iii} & W86, S97 & $-$2.461  \\
0336$-$019 & H$\beta$, Mg\,{\sc ii}, C\,{\sc iii} & B89, JB91 & $-$2.794  \\
0414$-$189 & Mg\,{\sc ii}, C\,{\sc iv}, C\,{\sc iii} & H78 & $+$0.026  \\
0420$-$014 & Mg\,{\sc ii}, C\,{\sc iii} & B89, S97 & $-$2.924  \\
0440$-$003 & Mg\,{\sc ii} & B89 & $-$2.069  \\
0454$-$234 & Mg\,{\sc ii}, C\,{\sc iii} & S89 & $-$1.872 \\
0454$-$463 & Mg\,{\sc ii} & F83 & $-$1.837  \\
0458$-$020 & C\,{\sc iv} & B89 & $-$1.702  \\
0521$-$365 & H$\beta$   & SK93 & $-$2.722 \\
0537$-$441 & Mg\,{\sc ii} & W86, S97 & $-$2.234  \\
0539$-$057 & Mg\,{\sc ii} & S93 & $-$1.390  \\
0804$+$449 & Mg\,{\sc ii}, C\,{\sc iv}, C\,{\sc iii} & L96& $+$0.411  \\
0836$+$710 &Mg\,{\sc ii}, C\,{\sc iv}, C\,{\sc iii}, Ly$\alpha$ &L96&$+$0.258 \\
0851$+$202 & H$\beta$   & S89 & $-$4.109  \\
0954$+$556 & C\,{\sc iv}, Ly$\alpha$ & W95 & $-$2.063 \\
0954$+$658 & H$\alpha$  & L96 & $-$4.425  \\
1127$-$145 & Mg\,{\sc ii}, C\,{\sc iv}, C\,{\sc iii} & W86, W95 & $-$0.317  \\
1222$+$216 & H$\beta$ & SM87 & $-$2.109  \\
1226$+$023 & H$\beta$, Mg\,{\sc ii}, C\,{\sc iv}, C\,{\sc iii},
Ly$\alpha$, H$\alpha$ & K00, U80, CB96, B87 & $-$2.067  \\
1229$-$021 & Mg\,{\sc ii} & B98 & $-$1.729 \\
1253$-$055 & H$\beta$, Mg\,{\sc ii}, C\,{\sc iv}, C\,{\sc iii},
Ly$\alpha$ & W95, M96 & $-$1.797  \\
1331$+$170 & C\,{\sc iv}  & C92  & $-$1.716  \\
1334$-$127 & Mg\,{\sc ii} & SK93 & $-$2.370  \\
1424$-$418 & Mg\,{\sc ii}, C\,{\sc iii} & S89 & $-$2.194  \\
1504$-$199 & Mg\,{\sc ii} & H78 & $-$1.898  \\
1510$-$089 & H$\beta$     & O94 & $-$2.595  \\
1611$+$343 & H$\beta$, C\,{\sc iv}, C\,{\sc iii}, Ly$\alpha$ & W95 & $-$3.075 \\
1633$+$382 & Mg\,{\sc ii}, C\,{\sc iv}, C\,{\sc iii}, Ly$\alpha$ &L96& $-$2.537 \\
1725$+$044 & H$\beta$     & R84 & $-$2.819  \\
1739$+$522 & Mg\,{\sc ii}, C\,{\sc iv}, C\,{\sc iii} &L96 & $-$1.446  \\
1741$-$038 & Mg\,{\sc ii} & S89 & $-$1.816  \\
2230$+$114 & C\,{\sc iv}, C\,{\sc iii}, Ly$\alpha$ & W95 & $-$1.856  \\
2251$+$158 & H$\beta$, C\,{\sc iv}, C\,{\sc iii}, Ly$\alpha$ &W95 & $-$2.632 \\
2320$-$035 & C\,{\sc iv} & B89 & $-$1.653 \\
2351$+$456 & Mg\,{\sc ii}, C\,{\sc iv}, C\,{\sc iii} & L96 & $-$1.480 \\
\enddata
\tablerefs{B75: \citet{B75}; B89: \citet{B89}; C92: \citet{C92};
CB96: \citet{CB96}; F83: \citet{F83}; G01: \citet{G01}; H78:
\citet{H78}; JB91: \citet{JB91}; K00: \citet{K00}; L96:
\citet{L96}; M96: \citet{M96}; O94: \citet{O94}; R84: \citet{R84};
S97: \citet{S97}; S89: \citet{S89}; S93: \citet{S93}; SK93:
\citet{SK93}; SM87: \citet{SM87}; U80: \citet{U80}; W86:
\citet{W86}; W95: \citet{W95}. }
\end{deluxetable}

\clearpage
\begin{deluxetable}{cccccccc}
\tabletypesize{\scriptsize} \tablecaption{$\gamma$-ray fluxes,
Doppler factors and core dominated parameters } \tablewidth{0pt}
\tablehead{ \colhead{Source} & \colhead{$\gamma$}&
\colhead{$F_{\gamma,\rm min}$} & \colhead{$F_{\gamma,\rm max}$} &
\colhead{$\delta _{\gamma,\rm min}$} & \colhead{$\delta
_{\gamma,\rm max}$} & \colhead{$\delta _{\rm var}$} &
\colhead{$R$} }\startdata
0119$+$041 & 2.63 & 5.1  & 20.3  & 5.98 & 7.23 & ...    & 15.85 \\
0208$-$512 & 1.99 & 35.2 & 134.1 & 7.76 & 9.71 & ...    & ...   \\
0336$-$019 & 1.84 & 37.4 & 177.6 & 7.27 & 9.56 & 19.01  & 31.62 \\
0414$-$189 & 3.25 & 10.2 & 49.5  & 7.63 & 9.19 & ...    & ...   \\
0420$-$014 & 2.44 & 9.3  & 64.2  & 6.61 & 8.75 & 11.72  & 251.3 \\
0440$-$003 & 2.37 & 12.5 & 85.9  & 8.52 & 11.3 & 11.46  & 19.95 \\
0454$-$234 & 3.14 & 8.1  & 14.7  & 5.94 & 6.39 & ...    & 5.00  \\
0454$-$463 & 2.75 & 5.5  & 22.8  & 7.15 & 8.65 & ...    & ...   \\
0458$-$020 & 2.45 & 9.5  & 68.2  & 10.1 & 13.5 & 17.80  & 5.01  \\
0521$-$365 & 2.63 & 15.8 & 31.9  & 4.46 & 4.91 & ...    & 1.00  \\
0537$-$441 & 2.41 & 16.5 & 91.1  & 9.03 & 11.6 & ...    & 199.6 \\
0539$-$057 & 2.00 & 66.5 & 66.5  & 4.60 & 4.60 & ...    & 7.61  \\
0804$+$499 & 2.15 & 8.3  & 15.1  & 3.82 & 4.20 & 26.21  & 151.9 \\
0836$+$710 & 2.62 & 8.6  & 33.4  & 3.19 & 3.85 & 10.67  & 34.93 \\
0851$+$202 & 2.03 & 9.7  & 15.8  & 9.62 & 10.4 & 18.03  & 781.0 \\
0954$+$556 & 2.12 & 6.5  & 47.2  & 3.44 & 4.73 & 4.63   & 10.81 \\
0954$+$658 & 2.08 & 6.6  & 18.0  & 14.5 & 17.0 & 6.62   & 110.5 \\
1127$-$145 & 2.70 & 9.9  & 61.8  & 4.87 & 6.24 & ...    & 50.12 \\
1222$+$216 & 2.28 & 6.9  & 48.1  & 7.12 & 9.57 & 8.16   & ...   \\
1226$+$023 & 2.58 & 8.5  & 48.3  & 3.04 & 3.87 & 5.71   & 6.31  \\
1229$-$021 & 2.85 & 4.9  & 15.5  & 5.45 & 6.33 & ...    & 3.43  \\
1253$-$055 & 1.96 & 9.3  & 267.3 & 5.18 & 9.14 & 16.77  & 12.59 \\
1331$+$170 & 2.41 & 4.4  & 33.1  & 6.20 & 8.34 & ...    & ...   \\
1334$-$127 & 2.62 & 5.5  & 20.2  & 4.71 & 5.64 & ...    & 12.87 \\
1424$-$418 & 2.13 & 15.3 & 55.3  & 5.28 & 6.48 & ...    & 2.38  \\
1504$-$166 & 2.00 & 16.5 & 33.2  & 5.58 & 6.27 & ...    & 19.95 \\
1510$-$089 & 2.47 & 12.6 & 49.4  & 5.84 & 7.11 & 13.18  & 31.62 \\
1611$+$343 & 2.42 & 19.0 & 68.9  & 6.36 & 7.68 & 5.04   & 25.12 \\
1633$+$382 & 2.15 & 31.8 & 107.5 & 5.22 & 6.33 & 8.83   & 79.44 \\
1725$+$044 & 2.67 & 13.3 & 23.7  & 8.68 & 9.39 & 2.46   & 125.8 \\
1739$+$522 & 2.42 & 9.7  & 44.9  & 6.72 & 8.40 & 12.12  & 50.1  \\
1741$-$038 & 2.42 & 11.7 & 48.7  & 4.44 & 5.47 & 8.92   & 4.00  \\
2230$+$114 & 2.45 & 12.1 & 51.6  & 5.92 & 7.31 & 14.23  & 25.12 \\
2251$+$158 & 2.21 & 24.6 & 116.1 & 5.97 & 7.61 & 21.84  & 15.85 \\
2320$-$035 & 2.00 & 8.2  & 38.2  & 12.1 & 15.6 & ...    & ...   \\
2351$+$456 & 2.38 & 11.8 & 42.8  & 6.37 & 7.71 & ...    & ...   \\
\enddata
\end{deluxetable}



\end{document}